\documentclass[twocolumn]{aastex63}
\shorttitle{Massive Black Hole Binary in NGC\,4472}
\shortauthors{Wrobel \& Lazio}

\begin{document}

\title{Astrometric Constraints on a Massive Black Hole Binary in NGC\,4472}

\author[0000-0001-9720-7398]{J. M. Wrobel}
\affiliation{National Radio Astronomy Observatory, P.O. Box O,
  Socorro, NM 87801, USA}

\author{T. J. W. Lazio}
\affiliation{Jet Propulsion Laboratory, California Institute of
  Technology, 4800 Oak Grove Drive, Pasadena, CA 91109, USA}

\correspondingauthor{J. M. Wrobel}
\email{jwrobel@nrao.edu}

\received{as ngVLA Memo \# 90}

\begin{abstract}
  In an EHT study of a Jy-level target, \citet{saf19} show how
  astrometric monitoring could constrain massive black hole binaries
  with the wide separations that make them long-lived against
  gravitational wave losses, and with the small mass ratios expected
  from merged satellite galaxies. With this ngVLA study, we show how
  such frontier topics could be explored for the more numerous
  mJy-level targets, such as NGC\,4472. We also discuss how ngVLA
  astrometric monitoring could test the upper limits from pulsar
  timing arrays on gravitational waves from NGC\,4472.
\end{abstract}


\section{Motivation}

The merger of two galaxies, each hosting a massive black hole (MBH),
could yield a bound MBH binary. Eventually the orbit of the MBH binary
will shink due to gravitational wave (GW) emission \citep{beg80}. Such
GW signatures are sought with pulsar timing arrays like the North
American Nanohertz Observatory for Gravitational Waves (NANOGrav), and
are projected to be detected with the future Laser Interferometer
Space Array. Searches are also underway for the electromagnetic (EM)
signatures of bound MBH binaries. Decades of data seeking indirect EM
signatures, such as periodicities in emission line velocities or
photometric variability, have been used to identify candidate MBH
binaries. See \citet{bur19} and \citet{der19} for recent reviews of
the aforementioned topics.

One direct EM signature would be to resolve and monitor the orbit of
one or both members of a bound MBH binary. EM strategies for achieving
this are only now being devised \citep{dor18,saf19,dex20}. These
strategies invoke recent or projected advances in interferometry at
millimeter or near-infrared wavelengths. Here, we consider a strategy
proposed by \citet{saf19} for continuum targets at Jansky (Jy) levels
observable at 230\,GHz (1.3\,mm) with the Event Horizon Telescope
\citep[EHT;][]{eht19}, and adapt it for the more numerous targets at
milliJansky (mJy) levels observable at 80\,GHz (3.7\,mm) with a
next-generation Very Large Array \citep[ngVLA;][]{mur18,sel18}.

\section{Target Selection}

We focus on NGC\,4472, Virgo Subcluster B's dominant galaxy which
shows evidence for size and mass growth via the accretion of satellite
galaxies \citep[e.g.,][]{cap15}. VLBI imaging of the low-luminosity
active galactic nucleus (LLAGN) in NGC\,4472 is available only at
5\,GHz and 8.4\,GHz \citep{nag05,and05}. The most constraining size
information comes from the VLBA image at 8.4\,GHz showing a 4-mJy
source \citep{and05}. At the assumed distance of 16.7\,Mpc
\citep[1\arcsec\, = 81.0\,pc;][]{bla09}, the upper limit on the radio
source's major axis, 730\,$\mu$as, corresponds to 0.059\,pc. Attempts
to probe smaller size scales via VLA time variability at 8.4\,GHz and
15\,GHz were inconclusive \citep{nag05,and05}. Hydrodynamical
simulations with radiative processes and MBH feedback suggest that the
LLAGN can be understood as jet-like emission launched from a
radiatively inefficient accretion flow \citep{ina20}.

\section{ngVLA Astrometric Observations}

We adopt the ngVLA's Long Baseline Array (LBA) at 80\,GHz, the
frequency assigned to notional continuum studies in the highest
frequency band \citep{wro20}. The resolution associated with the
maximum baseline has a point spread function $PSF =$ 90\,$\mu$as
(0.0073\,pc) at FWHM \citep{ros19}. To accrue a reasonable time on the
target, one antenna at each LBA station would observe the phase
calibrator while the rest continuously observed NGC\,4472
\citep[e.g.,][]{tho01}. After 10 hours on target the RMS noise is
projected to be about 3\,$\mu$Jy\,beam$^{-1}$ \citep{ros19}.

The astrometric accuracy of the ngVLA observation will have
contributions from a term due to the signal-to-noise ratio $S/N$ on
the target, $\sigma_{\rm s/n}$, and a term due the relative accuracy
achieved via phase referencing, $\sigma_{\rm pr}$. We discuss both in
turn.

Regarding $\sigma_{\rm s/n}$, an ALMA archival image at 98\,GHz
(3.1\,mm) with a $PSF =$ 0.61\arcsec\, (49\,pc) shows a 2-mJy source
that is slightly resolved, possibly due to residual phase calibration
errors. If only its peak signal of 1.3\,mJy\,beam$^{-1}$ is available
to the LBA at 80\,GHz, it would have an $S/N \sim 430$ and be
localized with an associated accuracy of $\sigma_{\rm s/n} = PSF /
(1.665 \times S/N) \sim 0.1\,\mu$as. Regarding $\sigma_{\rm pr}$, a
sub-degree separation between NGC\,4472 and its phase calibrator would
be required to reach $\sigma_{\rm pr} \sim 1\,\mu$as \citep{rei18}.
One such a calibrator is already known.

\begin{figure*}[!thb]
\centering
\includegraphics[angle=0,scale=0.9]{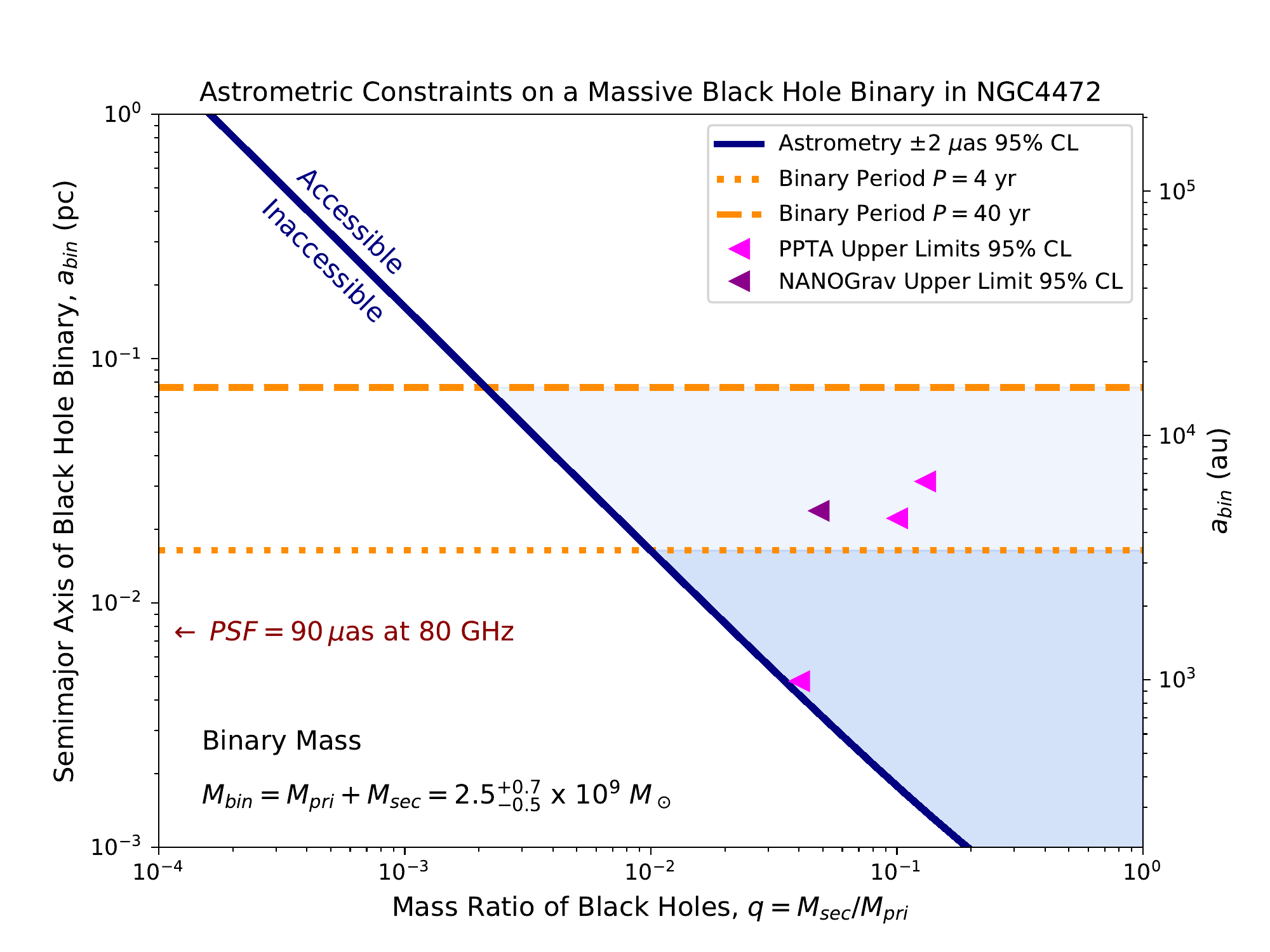}
\caption{Parameter space for $a_{bin}$ and $q$ for a hypothetical MBH
  binary in NGC\,4472, the dominant galaxy in Virgo Subcluster B.
  $M_{bin}$ is from \citet{rus13}. GW constraints from the PPTA are
  tabulated in \citet{sch16}, while that from NANOGrav is derived from
  \citet{agg19}.}
\end{figure*}

We are preparing a VLBA proposal on NGC\,4472 to help improve our
estimate of $\sigma_{\rm s/n} \sim 0.1\,\mu$as. For now, we assume
that the astrometric accuracy of the ngVLA observation will be
dominated by the $\sigma_{\rm pr} \sim 1\,\mu$as term.

\section{Interpretation}

\subsection{ngVLA Astrometric Monitoring}

Following \citet{saf19}, we examine how astrometric monitoring of
NGC\,4472 could contrain the reflex motion of the LLAGN's primary
black hole of mass $M_{pri}$, as it is tugged on by a putative
secondary black hole of mass $M_{sec}$. These masses define a binary
mass $M_{bin} = M_{pri} + M_{sec}$ and mass ratio $q = M_{sec} /
M_{pri} \le 1$. We allow $q$ to vary but adopt $M_{bin} =
2.5^{+0.7}_{-0.5} \times 10^9$\,M$_{\odot}$ from the total black-hole
mass estimated by \citet{rus13}.

We assume a circular orbit for the MBH binary and use Kepler's third
law to tie the binary's semimajor axis $a_{bin}$ to its mass $M_{bin}$
and orbital period $P$ \citep[equation (1) of][]{dex20}. The reflex
motion of $M_{pri}$ as it orbits, with semimajor axis $a_{pri}$, about
the binary's center of mass is $a_{pri} = a_{bin} \times q / (1+q)$
\citep[equation (1) of][]{saf19}. We recast this as

\begin{equation}
a_{bin} = a_{pri} \times (1+q) / q. \label{(1)}
\end{equation}

If astrometric monitoring of the LLAGN position could achieve a 95\%
accuracy of $2\,\mu$as, the reflex constraint would become $a_{pri} =
2\,\mu{\rm as} \times 81.0\,{\rm pc} / 10^6\,\mu{\rm as} =$
0.00016\,pc. Inserting this value for $a_{pri}$ into equation (1) then
defines how $a_{bin}$ varies with $q$. This behavior is shown as the
navy diagonal line in Figure~1. The parameter space to the right
(left) of this line is acessible (inaccessible) via astrometric
monitoring of NGC\,4472 with the adopted accuracy. The $PSF$ of the
ngVLA at 80\,GHz is marked for reference in Figure~1.

Figure~1 also indicates the values of $a_{bin}$ associated with two
fiducial MBH binary periods, $P = 4$\,yr and $P = 40$\,yr. Astrometric
monitoring through a quarter of a period would be sufficient to
constrain the range of $q$ values allowed for the period
\citep{saf19}. Therefore, if no reflex motion is detected for
$M_{pri}$ after 1 year of monitoring, then an MBH binary with $P =
4$\,yr and $q > 0.01$ could be excluded. The darker blue shading in
Figure~1 indicates where MBH binaries with shorter periods and higher
mass ratios might also be excluded.

If reflex motion remains undetected after a decade of astrometric
monitoring of NGC\,4472, then an MBH binary with $P = 40$\,yr and $q >
0.003$ could be excluded. Shorter periods with higher mass ratios
could also be excluded, as shown by the lighter blue shading in
Figure~1.

As \citet{saf19} highlight in their EHT study of a Jy-level target,
using astrometry to open a broad parameter space in $a_{bin}$ and $q$
enables probes of MBH binaries with the wider separations that make
them longer-lived against GW losses, and with the smaller mass ratios
expected from merged satellite galaxies. With this ngVLA study, we
have suggested how such frontier topics might be explored for the more
numerous mJy-level targets, such as NGC\,4472. Indeed, small $q$
values are expected for NGC\,4472, given independent evidence that it
has built up its size and mass by accreting satellite galaxies
\citep[e.g.,][]{cap15}.

In contrast to many previous efforts that aim to detect the radio
emission from a pair of MBHs in a merger remnant, this strategy
requires that only one of them have associated radio emission. And
importantly, the combination of the frequency coverage, angular
resolution, and sensitivity of the ngVLA LBA would enable searches
well into the regime at which an MBH binary is emitting GWs, unlike
previous efforts \citep[e.g.,][]{ban17}. We elaborate further on this
topic in the next subsection.

\subsection{ngVLA Tie-Ins to GW Findings}

ngVLA astrometric monitoring could serve to independently test the GW
upper limits on $q$ plotted in Figure~1 for NGC\,4472 from the Parkes
Pulsar Timing Array \citep[PPTA;][]{sch16} and from NANOGrav at its
highest sensitivity, occuring at a GW frequency of 9\,nHz
\citep{agg19}. Values for $q$ are calculated using equation (5) of
\citet{sch16} with the GW-inferred chirp masses and the binary mass
from \citet{rus13}.

The NANOGrav search involves a GW frequency range of 2.8\,nHz to
317.8\,nHz. This range corresponds to orbital periods of a putative
MBH binary from $P = 22.6$\,yr to $P = 0.2$\,yr, equivalent to $a_{\rm
  bin} = 11000$\,au to $a_{\rm bin} = 460$\,au for $M_{bin} = 2.5
\times 10^9$\,M$_{\odot}$ \citep{rus13}. GW-derived axes are so small
that they are often expressed not in pc, but in au. (An au axis is
provided in Figure~1.) The $PSF$ adopted for the ngVLA astrometry of
NGC\,4472 corresponds to 1500\,au, making it complementary to and
midway within the range of axes constrained by \citet{agg19}. Those GW
constraints degrade significantly below about 5\,nHz, due to the
11\,yr data span. Future observations, potentially including ngVLA
pulsar timing observations \citep{cha18}, will improve the NANOGrav
constraints, and extend them to lower GW frequencies or longer orbital
periods.

\acknowledgments

The NRAO is a facility of the National Science Foundation (NSF),
operated under cooperative agreement by Associated Universities,
Inc.\ (AUI). The ngVLA is a design and development project of the NSF
operated under cooperative agreement by AUI.

The NANOGrav project receives support from NSF Physics Frontiers
Center award number 1430284. Part of this research was carried out at
the Jet Propulsion Laboratory, California Institute of Technology,
under a contract with the National Aeronautics and Space
Administration.

This paper makes use of the following ALMA data: ADS/JAO.ALMA\#
2015.1.00926.S. ALMA is a partnership of ESO (representing its member
states), NSF (USA) and NINS (Japan), together with NRC (Canada), MOST
and ASIAA (Taiwan), and KASI (Republic of Korea), in cooperation with
the Republic of Chile. The Joint ALMA Observatory is operated by ESO,
AUI/NRAO and NAOJ.

\end{document}